\documentstyle[12pt,twoside]{article}

\begin{document}

\title{IS THE SEMI-CLASSICAL ANALYSIS\\
VALID FOR EXTREME BLACK HOLES?}

\author{F.G. Alvarenga\thanks{%
e-mail: flavio@cce.ufes.br}, A.B. Batista\thanks{%
e-mail: brasil@cce.ufes.br},
J. C. Fabris\thanks{%
e-mail: fabris@cce.ufes.br}, G.T. Marques\thanks{%
e-mail: tadaiesky@bol.com.br}\\
Departamento de F\'{\i}sica, Universidade Federal do Esp\'{\i}rito Santo, \\
CEP29060-900, Vit\'oria, Esp\'{\i}rito Santo, Brazil}
\maketitle

\begin{abstract}
The surface gravity for the extreme Reissner-Nordstr\"om black
hole is zero suggesting that it has a zero temperature. However,
the direct evaluation of the Bogolubov's coefficients, using the standard
semi-classical analysis, indicates
that the temperature of the extreme black hole is ill definite:
the Bogolubov's coefficients obtained by performing
the usual analysis of a collapsing model of a thin shell, and employing the
geometrical optical approximation, do not
obey the normalization conditions. We argue that the failure of the employement of
semi-classical analysis for the extreme black hole is due to the absence of
orthonormal quantum modes in the vicinity of the event horizon in this particular case.

\vspace{0.7cm}

PACS number(s): 04.62.+v., 04.70.Dy

\end{abstract}

\section{Introduction}

The possibility that a black hole may radiate with a planckian
spectrum was first pointed out in the seminal paper of Hawking
\cite{hawking}. A collapsing model for the formation of the black
hole was considered: a spherical mass distribution collapses under
the action of gravity, leading to a final state where all mass is
hidden behind an event horizon. The initial state is a Minkowski
space-time. The final state is the Schwarzschild space-time which
is asymptotically flat. Quantum fields are considered in this
dynamical configuration. The main point is that the initial vacuum
state does not coincide with the final vacuum state. From the
point of view of an observer at the spatial infinity, after the
formation of the black hole particles are created with a
planckian spectrum. This fact allows to attribute to the black
hole a temperature. For the Schwarzschild black hole, the
temperature is $T = 1/(8\pi M)$, where $M$ is the mass
of the black hole.
\par
The analysis of Hawking radiation for more general cases, like the
Reissner-Nordstr\"om black hole, leads also to the notion of
temperature due to the planckian form of the spectrum of the
emitted particles. In general, the expression for the temperature
for a black hole, using a semi-classical analysis (in the
sense of propagation of quantum fields in the geometric optical approximation), is given by $T
= \kappa/2\pi$ where $\kappa$ is the surface gravity. The Hawking
radiation is also well definite for rotating black holes, charged
or not. A particular case occurs for the so-called extreme black
holes, for which the surface gravity is zero. In this case, it is
generally stated that the temperature is also zero. In fact,
considering the Reissner-Nordstr\"om solution, the expression for
the temperature is given by
\begin{equation}
T = \frac{1}{8\pi M}\biggr(1 - 16\pi^2\frac{Q^4}{A^2}\biggl) \quad ,
\end{equation}
where $A = 4\pi r_+^2$, $r_+ = M + \sqrt{M^2 - Q^2}$, $Q$ is the
charge of the black hole and $M$ its mass. In the limit $Q = M$, $T = 0$.
\par
However, this definition of the temperature of the extreme black
hole as the limit of the temperature of the non-extreme black hole
when $Q \rightarrow M$ may hide some subtle points about the
thermodynamics of the extreme black hole. The main question we would like to address is
the following: is
it possible to obtain the zero temperature by performing a semi-classical
analysis if the extreme
condition is imposed from the beginning? The goal of this work is
to show that such analysis contains many controversial aspects and
it is very like that no semi-classical analysis is possible for
the extreme black hole.
\par
There has been many discussions on the real existence of an extreme black hole.
Perturbative considerations based on the expansion of the energy-momentum tensor of
quantum fields coupled to Einstein's equations led to doubts on the possibility to
have extreme black hole solutions \cite{anderson}. Moreover, it has been argued that
the existence of a zero temperature black hole would violate the third law of thermodynamics
unless the weak energy condition is not satisfied \cite{israel}. But, an analysis of the
collapse of a charged thin shell indicates that, classically, an extreme black hole can
be formed \cite{boul}. One assumption of the present paper, based on the results of
reference \cite{boul}, is that an extreme black hole can be formed through gravitational
collapse.
\par
Let us review the main steps of the evaluation of the temperature of a black
hole sketched above.
The semi-classical analysis of the thermodynamics of a black hole
is generally performed considering the formation of the black hole
due to the gravitational collapse of a spherical distribution of
mass (see references \cite{hawking,ford,spindel,birrel,wald}).
Initially the mass density is almost zero, and the
space-time is the Minkowski one. Later, the collapse of the mass
distribution leads to the formation of an event horizon, which
characterizes the black hole. The space-time after the formation
of the black hole is asymptotically flat. The
vacuum state of a quantum field before the collapse of the mass
distribution to a black hole does not coincide with the vacuum
state for the same quantum field after the appearance of the black
hole. The computation of the Bogolubov's coefficients between
the {\it in} and {\it out} quantum states leads to the notion of
temperature. The Bogolubov's coefficients, which allows to
express the quantum states {\it out} in terms of the quantum states
{\it in}, must obey some normalization conditions, which can also
be seen as a set of compatibility conditions.
\par
The application of the procedure described above to the extreme
Reissner-Nordstr\"om black hole has been discussed in references
\cite{liberati1,liberati2,gao}. In references \cite{liberati1,liberati2} it
has been pointed out that the extreme black hole does not behave
as thermal object. Moreover, the number of particle created for
each frequency $\omega$ is infinite. The authors have exploited an
analogy between the extreme black hole and the uniformly
accelerated mirror. These conclusions have been criticized in reference
\cite{gao} who argued that a modification in the calculations is
needed in order to give sense to some mathematical steps.
Moreover, the author of reference \cite{gao} has considered a wave packet instead of a simple
plane wave expansion.
\par
It is well known that the construction of a wave packet may eliminate
divergent expressions when a pure plane wave expansion is
considered. This possibility was already stressed in
\cite{liberati1}. In the present paper we would like to point out
that the problem of evaluation of the Bogolubov's coefficients
for the extreme black holes is more delicate than stated in
references \cite{liberati1,liberati2}. Not only the number of
particles are infinite, but also the normalization conditions for
the Bogolubov's coefficients are not obeyed. This is due essentially to
the
properties of the Bogolubov's coefficient $\alpha_{\omega\omega'}$:
in the extreme case, the computation of the modulus of this
coefficient leads to non-convergent integrals. This may indicate that the computation of
Bogolubov's coefficients has no sense for the extreme case, at least in the framework
of a semi-classical analysis.
\par
The modification introduced by \cite{gao} does not change the
situation. This modification consists in considering a logarithmic
term which is present in the expression for the tortoise radial coordinate
$r^*$. This logarithmic term is sub-dominant near the horizon.
However, as it will be shown later, the logarithmic term is not
necessary, since all mathematical expression has a sense when it
is not considered. Even when it is taken into account, the problem
remains essentially the same. Moreover, the construction of a wave packet can not
change the result since it would consist, for the extreme case, in
the superposition of modes that do not obey the normalization
condition.
\par
The reason for this curious result is not obvious. But, we will argue that the failure of
semi-classical analysis lies
on the causal structure of the space-time generated by an extreme
black hole: the {\it in} and {\it out} states would not be connected
due to the fact that near the event horizon the quantum modes do
not admit an orthonormal basis. The loss of normalization of the quantum
modes near the horizon is due to the fact that the near horizon geometry is
a portion of
the anti-deSitter space-time. 
It is important to remark that the existence
of a zero temperature black hole would imply the violation of the third law of thermodynamics.
From this point of view, the results reported in this paper may just re-state the
validity of the third law of thermodynamics,
at least in the context of the semi-classical analysis.
\par
This paper is organized as follows. In next section, we review the
computation of Hawking radiation for the Reissner-Nordstr\"om
black holes. We use the simplified scenario of a thin charged
collapsing shell, following the analysis presented in reference
\cite{ford}. This allows us to fix notation and some important
relations, as the normalization conditions for the Bogolubov's
coefficients. In section 3, we redone this analysis for the
extreme case, showing explicitly that the normalization conditions
are not satisfied. In section 4, the behaviour of quantum modes near the horizon
is discussed. We present our conclusions in section 5.

\section{Hawking radiation for a Reissner-Nordstr\"om Black Hole}

The Reissner-Nordstr\"om solution for a static spherically
symmetric space-time with a constant radial electric field is
given by
\begin{equation}
ds^2 = \biggr(1 - \frac{2M}{r} + \frac{Q^2}{r^2}\biggl)dt^2
- \biggr(1 - \frac{2M}{r} + \frac{Q^2}{r^2}\biggl)^{-1}dr^2 - r^2(d\theta^2 + \sin^2\theta d\phi^2) \quad ,
\end{equation}
where $M$ is the mass and $Q$ is the charge. This solution can be rewritten as
\begin{equation}
\label{metric}
ds^2 = \biggr(1 - \frac{r_-}{r}\biggl)\biggr(1 - \frac{r_+}{r}\biggl)dt^2
- \biggl\{\biggr(1 - \frac{r_-}{r}\biggl)\biggr(1 - \frac{r_+}{r}\biggl)\biggl\}^{-1}dr^2 -
r^2(d\theta^2 + \sin^2\theta d\phi^2) \quad ,
\end{equation}
where $r_\pm = M \pm \sqrt{M^2 - Q^2}$. Black hole solutions
implies $M \geq Q$, while naked singularities appear if $M <
Q$. The case $M = Q$ corresponds to the extreme black hole
solution. This form of the metric leads to the new coordinates
\begin{eqnarray}
u = t - r^* \quad , \quad v = t + r^* \quad , \\
\label{torto1}
r^* = r + \frac{r_+^2}{r_+ - r_-}\ln\biggr[\frac{r}{r_+} - 1\biggl] - \frac{r_-^2}{r_+ - r_-}\ln\biggr[\frac{r}{r_-} - 1\biggl] \quad .
\end{eqnarray}
The metric may be written in terms of these new coordinates as
\begin{equation}
ds^2 = \biggr(1 - \frac{r_-}{r}\biggl)\biggr(1 - \frac{r_+}{r}\biggl)du\,dv -
r^2(d\theta^2 + \sin^2\theta d\phi^2) \quad .
\end{equation}
From now on, a two-dimensional model will be considered, ignoring the angular
terms. For the quantum fields to be considered later, this is equivalent to consider the angular term with $l = 0$ (zero angular
momentum). But, the final results do not depend on this assumption, being valid for
general $l$.
\par
Let us consider the simplified model of a collapsing thin shell.
The collapse of a charged thin shell has been studied in details
in reference \cite{boul}. When $t \rightarrow - \infty$,
the density of the shell goes to zero and the space-time is flat. Hence, at
past infinity a scalar quantum field may be expanded into the normal modes
\begin{equation}
\label{in}
\phi = \int d\omega\frac{1}{\sqrt{4\pi\omega}}\biggr(a_\omega e^{-i\omega v} + a^\dag_\omega e^{i\omega v}\biggl) \quad ,
\end{equation}
where we have just considered the incoming modes given by the
coordinate $v$. After the collapse of the shell, a black hole is
formed given a space-time described by the metric (\ref{metric}),
which is asymptotically flat. Hence the outcome mode at $t
\rightarrow \infty$ is given by
\begin{equation}
\label{out}
\phi = \int d\omega\frac{1}{\sqrt{4\pi\omega}}\biggr(b_\omega e^{-i\omega u} + b^\dag_\omega e^{i\omega u}\biggl) \quad .
\end{equation}
The problem to solve is how to connect the coordinates $u$ and $v$, obtaining in this way
the Bogolubov's coefficients of the transformation
\begin{equation}
\frac{e^{-i\omega u}}{\sqrt{4\pi\omega}} = \int_0^\infty\biggr\{\alpha_{\omega\omega'}e^{-i\omega'v} +
\beta_{\omega\omega'}e^{i\omega'v}\biggl\}\frac{d\omega'}{\sqrt{4\pi\omega'}}
\end{equation}
with the inverse transformation
\begin{equation}
\frac{e^{-i\omega v}}{\sqrt{4\pi\omega}} = \int_0^\infty\biggr\{\alpha^*_{\omega'\omega}e^{-i\omega'u} -
\beta_{\omega'\omega}e^{i\omega'u}\biggl\}\frac{d\omega'}{\sqrt{4\pi\omega'}} \quad .
\end{equation}
The coefficient $\beta_{\omega\omega'}$ is connected with the number of particles
detected by an observer in the future infinity for each frequency $\omega$.
The Bogolubov's coefficients satisfy the consistency relations,
\begin{eqnarray}
\label{c1}
\int_0^\infty\biggr\{\alpha_{\omega\omega'}\alpha^*_{\omega''\omega'} -
\beta_{\omega\omega'}\beta^*_{\omega''\omega'}\biggl\}d\omega' &=& \delta(\omega - \omega'')
\quad ,\\
\label{c2}
\int_0^\infty\biggr\{\alpha_{\omega\omega'}\beta_{\omega''\omega'} -
\beta_{\omega\omega'}\alpha_{\omega''\omega'}\biggl\}d\omega' &=& 0 \quad .
\end{eqnarray}
Notice that
\begin{equation}
\label{c3}
\int_0^\infty\int_0^\infty\biggr\{\alpha_{\omega\omega'}\alpha^*_{\omega''\omega'} -
\beta_{\omega\omega'}\beta^*_{\omega''\omega'}\biggl\}d\omega'd\omega'' = 1 \quad ,
\end{equation}
a normalization condition to be used later.
\par
An incoming mode $v$ comes from the infinity past, traversing the
collapsing shell, becoming later an outcome mode, traversing again the shell and attaining
the infinity future. The modes are continuous, in the sense that
$v|_{R=R_1} = V|_{R=R_1}$, $V|_{R = 0} = U|_{R = 0}$, $U|_{R=R_2}
= u|_{R=R_2}$, where $u$ and $v$ are the outgoing and incoming
modes in the external geometry determined by the shell, while $U$
and $V$ are the same modes in the internal, minkowskian geometry,
and $R_1$ and $R_2$ are the radius of the shell at the first and
second crossing, respectively. An important point in this
derivation is that the collapse is accelerated in such a way that
the speed of the collapsing shell approach the speed of the light
at the moment when the event horizon is formed.
\par
At the second crossing, which we admit to occur near the moment of formation of the
black hole, the continuity of the metric leads to
\begin{equation}
\label{col}
dT^2 - dR^2 = \biggr[1 - \frac{r_+}{R}\biggl]\biggr[1 - \frac{r_-}{R}\biggl]dt^2 -
\biggr\{\biggr[1 - \frac{r_+}{R}\biggl]\biggr[1 - \frac{r_-}{R}\biggl]\biggl\}^{-1}dR^2 \quad .
\end{equation}
At the moment of the second crossing, we may consider that
\begin{equation}
\label{colapso}
R \approx r_+ + A(T_0 - T)\quad ,
\end{equation} where $A$ is a constant
and $T_0$ is the time where the black hole is formed. Inserting
this relation in (\ref{col}), and considering the continuity of
the incoming modes, what allows to express $T$ in terms of $v$, it
results the following relation between the $u$ and $v$ modes:
\begin{equation}
\label{rela}
u = - 2\sigma\ln\biggr[\frac{v_0 - v}{C}\biggl]
\end{equation}
where $\sigma = \kappa^{-1} = r_+^2/(r_+ - r_-)$ is the inverse of the surface gravity, $C$ is a constant and
$v_0 = T_0 - r_+$.
Using the inner product for complex scalar fields,
\begin{equation}
(\phi_1,\phi_2) = - i\int d\Sigma^\mu\biggr(\phi_1\partial_\mu\phi^*_2 - \partial_\mu\phi_1\phi_2^*\biggl) = -i\int d\Sigma^\mu\phi_1\stackrel{\leftrightarrow}{\partial}_\mu\phi_2^* \quad ,
\end{equation}
and defining $f_\omega = e^{-i\omega v}/\sqrt{4\pi\omega}$, $g_\omega = e^{-i\omega u(v)}/\sqrt{4\pi\omega}$, we obtain the
following expressions for the Bogolubov coefficients:
\begin{equation}
\label{coef}
\alpha_{\omega\omega'} = (g_\omega,f_{\omega'}) \quad , \quad \beta_{\omega\omega'} = - (g_\omega,f^*_{\omega'}) \quad .
\end{equation}
Due to the relation between $u$ and $v$ (\ref{rela}), the Bogolubov coefficients can be expressed in terms of the integrals
\begin{eqnarray}
\alpha_{\omega\omega'} = \frac{1}{4\pi\sqrt{\omega\omega'}}\exp[i(\omega' v_0 -
2\sigma\omega\ln C - 2\sigma\omega\ln\omega')]\int_0^\infty e^{-i(y - 2\sigma\omega\ln y)}
\biggr\{1 + \frac{2\sigma\omega}{y}\biggl\}dy \quad , \\
\beta_{\omega\omega'} = \frac{1}{4\pi\sqrt{\omega\omega'}}\exp[-i(\omega' v_0 +
2\sigma\omega\ln C + 2\sigma\omega\ln\omega')]\int_0^\infty e^{i(y + 2\sigma\omega\ln y)}
\biggr\{1 - \frac{2\sigma\omega}{y}\biggl\}dy \quad ,
\end{eqnarray}
where $y = \omega'(v_0 - v)$. Making an integration by parts and
using the integral relation \footnote{It could be argued that this expression is not
definite strictly speaking when the exponential is pure
imaginary. However, the limit case of a pure imaginary
exponential term is well definite.}
\begin{equation}
\int_0^\infty e^{\pm iy}y^{ia}dy = \frac{\Gamma[1 + ia]}{(\pm i)^{\mp i\pi/2 + ia}} = \mp
a\Gamma(ia)e^{\mp\pi/2} ,
\end{equation}
it results the following expressions for the Bogolubov's coefficients:
\begin{eqnarray}
\alpha_{\omega\omega'} &=& \frac{2\sigma\omega}{2\pi\sqrt{\omega\omega'}}\Gamma(2i\sigma\omega)
e^{\sigma\pi\omega}\exp[i(\omega'v_0 - 2\sigma\omega\ln C - 2\sigma\omega\ln\omega')] \quad ,\\
\beta_{\omega\omega'} &=& - \frac{2\sigma\omega}{2\pi\sqrt{\omega\omega'}}\Gamma(2i\sigma\omega)
e^{-\sigma\pi\omega}\exp[-i(\omega'v_0 + 2\sigma\omega\ln C + 2\sigma\omega\ln\omega')] \quad .
\end{eqnarray}
It is important to stress that imposing the extreme condition $\sigma \rightarrow \infty$,
the $\beta$ coefficient becomes zero.
\par
Before evaluating the Hawking temperature, we must notice that the above solutions
for the Bogolubov's coefficients satisfies the consistency relations (\ref{c1},\ref{c2}).
It will be shown only the relation (\ref{c1}) rewritten as in (\ref{c3}). Computing the
first term in the left hand side of (\ref{c3}), it comes out,
\begin{eqnarray}
& &\int_0^\infty\int_0^\infty d\omega'd\omega''\alpha_{\omega\omega'}\alpha^*_{\omega''\omega'} \nonumber\\
&=&
\frac{\sigma^2}{\pi^2}\int_0^\infty\int_0^\infty\frac{d\omega'}{\omega'}d\omega''
\sqrt{\omega\omega''}\Gamma(2i\sigma\omega)\Gamma(-2i\sigma\omega'')e^{\pi\sigma(\omega + \omega'')} e^{\{-2i\sigma(\omega - \omega'')[\ln C + \ln\omega']\}} \quad, \nonumber\\
&=& \frac{2\sigma^2}{\pi}\int_0^\infty d\omega''\sqrt{\omega\omega''}\Gamma(2i\sigma\omega)
\Gamma(-2i\sigma\omega'')e^{\pi\sigma(\omega + \omega'')}
e^{[-2i\sigma(\omega - \omega'')]}\delta[2\sigma(\omega - \omega'')] \quad ,\nonumber \\
&=&\frac{1}{2}\frac{e^{2\pi\sigma\omega}}{\sinh(2\pi\sigma\omega)} \quad ,
\end{eqnarray}
where we have used the relation
\begin{equation}
\Gamma(ia)\Gamma(-ia) = \frac{\pi}{a\sinh(\pi a)} \quad .
\end{equation}
A similar calculation leads to
\begin{equation}
\int_0^\infty\int_0^\infty d\omega'd\omega''\beta_{\omega\omega'}\beta^*_{\omega''\omega'}
= \frac{1}{2}\frac{e^{-2\pi\sigma\omega}}{\sinh(2\pi\sigma\omega)} \quad .
\end{equation}
From these expressions, the normalization condition (\ref{c3}) can be easily obtained:
\begin{eqnarray}
\int_0^\infty\int_0^\infty d\omega'd\omega''\alpha_{\omega\omega'}\alpha^*_{\omega''\omega'} &-&
\int_0^\infty\int_0^\infty d\omega'd\omega''\beta_{\omega\omega'}\beta^*_{\omega''\omega'} =
\nonumber\\
&=& \frac{1}{2}\frac{e^{2\pi\sigma\omega}}{\sinh(2\pi\sigma\omega)}
- \frac{1}{2}\frac{e^{-2\pi\sigma\omega}}{\sinh(2\pi\sigma\omega)} = 1 \quad .
\end{eqnarray}
\par
The Hawking temperature can be obtained in two equivalent ways.
First by computing the number of particles with frequency $\omega$
in the future infinity,
\begin{equation}
N_\omega = \int_0^\infty\int_0^\infty d\omega'd\omega''\beta_{\omega\omega'}\beta^*_{\omega''\omega'}
\end{equation}
or by noticing that
\begin{equation}
||\alpha_{\omega\omega'}|| = e^{\pi\sigma\omega}||\beta_{\omega\omega'}||
\end{equation}
and using the normalization condition (\ref{c3}). In both cases the result is
\begin{equation}
N_\omega = \frac{1}{e^{2\pi\sigma\omega} - 1} \quad .
\end{equation}
This is characteristic of a Planckian spectrum with temperature $T
= 1/(2\pi\sigma)$. In the non-extreme Reissner-Nordstr\"om case
treated before, this temperature reads
\begin{equation}
T = \frac{1}{8\pi M}\biggr(1 - \frac{16\pi^2Q^4}{A^2}\biggl) \quad ,
\end{equation}
where $A = 4\pi r_+^2$ is the area of the event horizon. It can be verified that
when $Q \rightarrow M$, $T \rightarrow 0$.

\section{The extreme black hole}

The extreme condition $M = Q$, leads to the metric
\begin{equation}
\label{em}
ds^2 = \biggr\{1 - \frac{M}{r}\biggl\}^2dt^2 - \biggr\{1 - \frac{M}{r}\biggl\}^{-2}dr^2 -
r^2(d\theta^2 + \sin^2\theta d\phi^2) \quad .
\end{equation}
The null coordinates take now the form
\begin{eqnarray}
u = t - r^* \quad , \quad v = t + r^* \quad ,\\
\label{torto2}
r^* = r + 2M\ln\biggr(\frac{r}{M} - 1\biggl) - \frac{M}{r/M - 1} \quad ,
\end{eqnarray}
The extreme black hole has a degenerate event horizon with $r_+ =
r_-$. The new tortoise coordinate (\ref{torto2}) is quite different from
the preceding one for the non-extreme case, (\ref{torto1}), mainly due to
the last term in (\ref{torto2}). But, as it can be verified, expression (\ref{torto2})
may be obtained as a limit case of (\ref{torto1}) when $r_- \rightarrow r_+$.
\par
The same model of the preceding section will be considered now on:
a collapsing thin shell, with the space-time external to the shell
being determined by the metric (\ref{em}), while the internal
space-time is minkowskian. At the past infinity, all space-time is
essentially minkowskian and a quantum scalar field admits the
decomposition (\ref{in}), while in the future infinity the
space-time is asymptotically minkowskian and the quantum scalar field admits
the decomposition (\ref{out}). The task now is the same as before:
to connect both quantum states.
\par
In this sense, an ingoing mode, described by the null coordinate
$v$ comes from the past infinity, traverses the thin shell when space-time
is essentially Minkowski, becoming later an outgoing
mode, which traverses again the shell near the moment of the
formation of the black hole. The same match conditions established
before can be used. Repeating all the calculations performed
before we find now
\begin{equation}
\label{relb}
u = \frac{C}{v_0 - v} \quad ,
\end{equation}
where, as before, $C$ and $v_0$ are constants connected with some parameters characterizing
the collapse of the shell. In performing this evaluation, we considered just the leading
term near the horizon in (\ref{torto2}). The problem of taking into account the sub-dominant
logarithmic term will be discussed later. But, we may already state that taking into
account this sub-dominant term does not change the essential of the results.
\par
The expressions (\ref{coef}) are used again to compute the Bogolubov coefficients, leading
to
\begin{eqnarray}
\alpha_{\omega\omega'} &=& (g_\omega,f_{\omega'}) = \frac{1}{4\pi\sqrt{\omega\omega'}}
\int_{-\infty}^{v_0}e^{-i\frac{\omega C}{v_0 - v} + i\omega'v}\biggr\{\omega' + \frac{\omega C}{(v_0 - v)^2}\biggl\}dv \nonumber \quad  \\
&=& \frac{e^{i\omega'v_0}}{4\pi\sqrt{\omega\omega'}}\int_0^\infty e^{-i(y + D/y)}\biggr\{1 + \frac{D}{y^2}\biggl\}dy \quad ; \\
\beta_{\omega\omega'} &=& - (g_\omega,f^*_{\omega'}) = \frac{1}{4\pi\sqrt{\omega\omega'}}
\int_{-\infty}^{v_0}e^{-i\frac{\omega C}{v_0 - v} - i\omega'v}\biggr\{\omega' - \frac{\omega C}{(v_0 - v)^2}\biggl\}dv \nonumber \quad  \\
&=& \frac{e^{-i\omega'v_0}}{4\pi\sqrt{\omega\omega'}}\int_0^\infty e^{i(y - D/y)}\biggr\{1 - \frac{D}{y^2}\biggl\}dy \quad ,
\end{eqnarray}
with $y = \omega'(v_0 - v)$. With the aid of the variable
redefinition $y = \pm D/z$, $D = C\omega'\omega$, the expression
for the Bogolubov's coefficients take the form,
\begin{eqnarray}
\label{alpha1}
\alpha_{\omega\omega'} &=& \frac{1}{2\pi}\frac{e^{i\omega'v_0}}{\sqrt{\omega\omega'}}D
\int_0^\infty e^{-i(y + D/y)}\frac{dy}{y^2} \quad , \\
\label{beta1}
\beta_{\omega\omega'} &=& \frac{1}{4\pi}\frac{e^{-i\omega'v_0}}{\sqrt{\omega\omega'}}D
\biggr\{\int_{-\infty}^0e^{i(y - D/y)}\frac{dy}{y^2} - \int_0^\infty e^{i(y - D/y)}\frac{dy}{y^2}\biggl\} \quad .
\end{eqnarray}
These integrals may be solved by using the integral representation of modified Bessel
functions of second kind,
\begin{equation}
\label{ir}
K_\nu(z) = \frac{1}{2}\biggr(\frac{z}{2}\biggl)^\nu\int_0^\infty\frac{e^{-t - \frac{z^2}{4t}}}{t^{\nu + 1}}dt \quad .
\end{equation}
Using the expressions for the Bogolubov's coefficients
(\ref{alpha1},\ref{beta1}) and this integral representation, it
results \footnote{Notice that the integral representation
(\ref{ir}) is valid for $|\arg z| < \pi/2$. The computation of
$\alpha_{\omega\omega'}$ implies $|\arg z| = \pi/2$. This limit
case is, however, well definite. This can be seen by expanding the
exponential in terms of sinus and cosinus and using the integrals
$3.868, 1-4$ of the Gradstein and Ryzhik table \cite{grad} after
differentiating them with respect to one of the parameters.}
\begin{eqnarray}
\label{bc1}
\alpha_{\omega\omega'} &=& \pm \frac{e^{i\omega'v_0}}{\pi}\sqrt{\frac{D}{\omega\omega'}}K_1(\pm 2i\sqrt{D}) = \mp \frac{e^{i\omega'v_0}}{\pi}\sqrt{\frac{D}{\omega\omega'}}H_1^{(1)}(\mp 2\sqrt{D})\quad , \\
\label{bc2}
\beta_{\omega\omega'} &=& \pm i\frac{e^{-i\omega'v_0}}{\pi}\sqrt{\frac{D}{\omega\omega'}}K_1(\pm 2\sqrt{D}) \quad ,
\end{eqnarray}
where $H_1^{(1)}(x)$ is the Hankel's function of first kind.
These expressions are very similar to the corresponding ones for
the uniformly accelerated mirror \cite{birrel}. Aside some
unimportant differences in constant factors, there is an
additional phase in the accelerated mirror problem, what can
distinguish crucially the mirror problem from the extreme black
hole problem.
\par
The first thing to remark about these expressions for the
Bogolubov coefficients, it is the absence of a simple relation
between them, in contrast of what happens with the non-extreme
case. This does not allow to extract the notion of a temperature
for the extreme case. This induces to state that the extreme black
hole is not a thermal object. However, the situation is more
complex yet. In fact, the Bogolubov coefficients found before do
not satisfy the normalization condition (\ref{c3}). Let us verify
now this, calculating separately the two terms of (\ref{c3}).
\par
Let us first compute the second term of (\ref{c3}). Using the solution found before, the
second term of (\ref{c3}) may be written as
\begin{equation}
\int_0^\infty\int_0^\infty d\omega'd\omega''\beta_{\omega\omega'}\beta^*_{\omega''\omega'} =
\frac{C}{\pi^2}\int_0^\infty\int_0^\infty d\omega'd\omega''K_1(2\sqrt{C\omega\omega'})K_1(2\sqrt{C\omega''\omega'}) \quad .
\end{equation}
The integral in $\omega''$ may be easily computed. In fact, it takes the form
\begin{equation}
\int_0^\infty d\omega''K_1(2\sqrt{C\omega''\omega'}) = \frac{1}{2C\omega'}\int_0^\infty dy\,y\,K_1(y)\end{equation}
where $y = 2\sqrt{C\omega'\omega''}$. In order to evaluate this integral, we use
the following integral representation of the modified Bessel function of second kind:
\begin{equation}
\label{ir2}
K_1(z) = \int_0^\infty e^{-z\cosh\theta}\cosh\theta\,d\theta \quad .
\end{equation}
Hence,
\begin{eqnarray}
\int_0^\infty K_1(y)y\,dy &=& \int_0^\infty\int_0^\infty e^{-y\cosh\theta}y\,\cosh\theta\,dyd\theta \nonumber\\
&=& \int_0^\infty\int_0^\infty e^{-x}x\,dx\,\frac{d\theta}{\cosh\theta} =
\int_0^\infty \frac{d\theta}{\cosh\theta} = \frac{\pi}{2} \quad ,
\end{eqnarray}
where we made the substitution $x = y\cosh\theta$.
In this way, the second term of (\ref{c3}) can be written as
\begin{eqnarray}
\int_0^\infty\int_0^\infty d\omega'd\omega''\beta_{\omega\omega'}\beta^*_{\omega''\omega'} &=&
\frac{1}{4\pi}\int_0^\infty K_1(2\sqrt{C\omega\omega'})\,\frac{d\omega'}{\omega'} \nonumber\\
&=&
- \frac{1}{2\pi}\int_0^\infty K_1(x)\,\frac{dx}{x} \nonumber\\
&=& -\frac{1}{2\pi}\int_0^\infty\int_0^\infty e^{-x\cosh\theta}\cosh\theta\,d\theta\,\frac{dx}{x} \quad .
\end{eqnarray}
where the integral representation of the function $K_1(x)$ (\ref{ir2}) has been used again.
Under the substitution $u = x\cosh\theta$, this term takes the form
\begin{equation}
\int_0^\infty\int_0^\infty d\omega'd\omega''\beta_{\omega\omega'}\beta^*_{\omega''\omega'} = \frac{1}{2\pi}\int_0^\infty\cosh\theta\,d\theta\int_0^\infty e^{-u}\,\frac{du}{u}
\end{equation}
which is obviously a divergent term.
\par
Until this moment, we may keep the hope to recover the
normalization condition, since this divergent term may be
cancelled by another divergent term coming from the first term in
(\ref{c3}). Let us now compute this term, choosing the minus sign
in (\ref{bc1}). It is more convenient to use the representation in terms
of Hankel's function in the expression for $\alpha_{\omega\omega'}$. Defining $x = 2\sqrt{C\omega''\omega'}$ and $y = 2\sqrt{C\omega\omega'}$, the final expression for the first term in (\ref{c3}) takes the form
\begin{equation}
\label{ff}
\int_0^\infty d\omega'd\omega''\alpha_{\omega\omega'}\alpha^*_{\omega''\omega'} = \frac{1}{4}\int_0^\infty H^{(2)}_1(x)x\,dx\int_0^\infty H_1^{(1)}(y)\,\frac{dy}{y} \quad ,
\end{equation}
where we have used the fact that, for real values
of the argument ${H^{(1)}}^*_1(x) = H^{(2)}_1(x)$, $H^{(2)}_1(x)$ being the Hankel's functions
of the second kind. The second integral in the right hand side of
(\ref{ff}) is a divergent term; however, the first integral is not
convergent, as an asymptotic analysis indicates. In fact, for large values of
the argument, $H_1^{(1,2)}(x) \rightarrow \sqrt{\frac{2}{\pi x}}e^{\pm(x - \frac{\pi}{2} -
\frac{\pi}{4})}$ and the integrand oscillates with increasing amplitude. The order
the
integrations are performed does not change this result
\par
With the results obtained above, it is easy to see that the
normalization condition is not satisfied. The second term in
(\ref{c3}) is infinite. However, the first term contains
non-convergent integrals; hence, its value is not definite. The
conclusion is that the Bogolubov transformation between the {\it in}
and {\it out} vacuum modes is ill definite for the extreme black hole,
and no thermodynamics can be constructed with a collapse scenario.
\par
We would like to stress that taking into account the logarithmic term
in (\ref{torto2}) does not change the scenario described before.
To take into account this term means to consider an expansion until second order.
In fact, it can be shown that it is equivalent to express the collapse of
the shell near the formation of the horizon as
\begin{equation}
R \approx M + A(T_0 - T) + B(T_0 - T)^2 \quad ,
\end{equation}
where $B$ is another positive parameter. This form for the collapse is still
consistent with the general analysis performed in reference \cite{boul}.
All the expansions must now be carried out
until second order, and the final results for the Bogolubov's coefficients are
now expressed as a sum of modified Bessel functions. However, the problems pointed out
above remain the same.

\section{Quantum fields near the horizon}

If a complete basis of orthonormal modes is implemented in the past infinity, the propagation
of the modes to the future infinity should in principle be well definite. The fact that
the normalization conditions are not satisfied when the Bogolubov's coefficients are
evaluated in the future infinity reveals that an anomaly occurs in the propagation of
quantum fields. Normally, the propagation of quantum fields is such that in any
hypersurface at constant time a complete basis can be definite. That is what occurs
in the non-extreme case. If the extreme case is the limit of the non-extreme one
when $Q \rightarrow M$ we could expect the same normal behaviour in the propagation of
quantum fields. 
\par
One important point is that the extreme RN black hole is not the limit
of the non-extreme one.
This has already been remarked in reference \cite{hawking1}. In this reference, the entropy law
for the extreme black hole has been studied, and the authors have concluded that
$S = 0$, in spite of the fact that the horizon area of the extreme black hole is non-zero.
Hence, a violation of the black hole entropy law $S = A/4$ occurs for the extreme RN black hole.
This results has been confirmed by the computation done in references \cite{mitra,zaslavskii}.
However, it has been argued that considering the string framework the usual entropy law
could be recovered \cite{horowitz}. This controversy shows that the thermodynamics of
extreme black holes is far from being a trivial subject.
\par
In reference \cite{hawking1} the problem of computation of the temperature of extreme black hole has been addressed by
employing the method of euclideanization of the metric. The conclusion was that the periodicity
of the euclidean time is arbitrary for the extreme black hole; hence the temperature is also
arbitrary and in any case non-zero. This result contrasts with the same evaluation
made for the non-extreme case, where a precise temperature can be obtained through the
periodicity of the euclidean time; moreover, the so-obtained temperature agrees with
the results obtained by using the surface gravity and by using the Bogolubov's coefficients.
This fact points again to the specificity of the extreme black hole, indicating that
it is not the limit case of the non-extreme black holes. In reference \cite{hawking1} this
specificity is connected with an unusual feature of the extreme black hole: as it can
be explicitly verified from the metric (\ref{em}), the spatial distance of any point
to the event horizon is infinite; for the non-extreme case, this spatial distance is finite.
\par
Another specific feature of the extreme black hole with respect to the non-extreme one concerns
the geometry near the event horizon. This specificity has a close connection with the
spatial distance to the event horizon quoted before.
For the non-extreme black hole, the metric near the horizon takes the form
\begin{equation}
\label{metric3}
ds^2 \approx \frac{r_+ - r_-}{r_+^2}\rho dt^2 - \frac{r_+^2}{r_+ - r_-}\frac{d\rho^2}{\rho} -
r_+^2d\Omega = \rho'dt'^2 - \frac{d\rho'^2}{\rho'} -r_+^2d\Omega^2 \quad ,
\end{equation}
where
\begin{equation}
\rho = r - r_+ \quad , \quad \rho'= \frac{r_+^2}{r_+ - r_-}\rho \quad , \quad t'=
\frac{r_+ - r_-}{r_+^2}t \quad .
\end{equation}
For the extreme black hole, however, the same computation leads to
\begin{equation}
\label{metric4}
ds^2 \approx \rho^2dt^2 - \frac{d\rho^2}{\rho^2} - M^2d\Omega^2 \quad .
\end{equation}
First of all, it must be remarked that the metric (\ref{metric4}) is not the limit of
(\ref{metric3}) when $r_- \rightarrow r_+$. Notice that the extreme limit and
the near horizon limit when applied for the metric (\ref{metric}) do not commute.
The specificity of the extreme black hole
with respect to the non-extreme ones comes from the geometry near the horizon. Far from
the horizon, the extreme space-time can be obtained from the non-extreme one.
\par
The geometry described by (\ref{metric4}) corresponds to an anti-deSitter space-time. More
precisely, to a portion of the anti-deSitter space-time. It is well known that the anti-deSitter
space-time has special problems concerning the propagation of initial data defined on a given
hyspersurface. In fact, the values of a given set of fields on a given hypersurface can be
obtained from the initial data of these fields on another hypersurface only in a portion
of the entire space-time. This is due to the fact that the anti-deSitter space-time has
a timelike infinity or, in other words, this space-time is not globally hyperbolic.
For a review of the properties of the anti-deSitter space-time, see \cite{ellis,petersen,douglas}. The formulation of a quantum field theory in an anti-deSitter
space-time has been studied in reference \cite{isham}. In this work, it has been shown that
no Hilbert space can be implemented in such space-time, unless specific boundary conditions
are fixed at infinity. To do so, it is necessary to use a universal covering of the anti-deSitter
space-time. However, the geometry described by the metric (\ref{metric4}) does not correspond
to this universal covering.
\par
From this considerations, we must expect that the problems with the normalization of
the Bogolubov's coefficients must come from the impossibility of assign a Hilbert space
for quantum fields in the space-time described by (\ref{metric4}). In fact, let us
solve the Klein-Gordon equation for the space-time described first by (\ref{metric3}) and
later by (\ref{metric4}). The quantum modes can be obtained from the classical solutions
with the canonical methods.
\par
Using (\ref{metric3}), the Klein-Gordon equation for a massless scalar field reduces to
\begin{equation}
\Box\phi = - \rho\phi'' - \phi' - \frac{\omega^2}{\rho} + \frac{l(l + 1)}{r_+^2}\phi = 0
\end{equation}
where primes denote derivatives with respect to $\rho$, $l$ is the angular momentum
eigenvalue and $\omega$ is normal mode frequency.
Solving this equation, we obtain
\begin{equation}
\phi = c_1K_{i4\omega}\biggr(\frac{\sqrt{l(l+1)}}{r_+}\sqrt{\rho}\biggl)e^{i\omega t} \quad ,
\end{equation}
where $K$ stands for the modified Bessel's function.
This is the well-known basis of orthonormal modes of a massless scalar field for RN and Schwarzschild
black holes (see \cite{matsas} and references therein). 
\par
For the extreme case, the employement of the (\ref{metric4}) reduces the Klein-Gordon
equation
to
\begin{equation}
- \rho^2\phi'' - 2\rho\phi' - \frac{\omega^2}{\rho^2} + l(l + 1)\phi = 0 
\quad ,
\end{equation}
with the solution
\begin{equation}
\label{kg1}
\phi = \frac{1}{\sqrt{\rho}}\biggr[c_1J_{l + 1/2}\biggr(\frac{\omega}{\rho}\biggl)
+\,c_2J_{-(l + 1/2)}\biggr(\frac{\omega}{\rho}\biggl)\biggl]e^{i\omega t}
\quad .
\end{equation}
Now, if we compute the norm of these modes, in the Klein-Gordon sense,
we obtain 
\begin{eqnarray}
\label{inner}
(\phi,\phi) &=& -i\int d\Sigma^\mu\phi\stackrel{\leftrightarrow}{\partial}_\mu\phi^*
 = 2\omega|c_{1,2}|^2\int_0^\infty \biggr[J_{\pm(l+1/2)}\biggr(\frac{\omega}{\rho}\biggl)\biggl]^2\frac{d\rho}{\rho^3}\nonumber\\
&=& 2\omega|c_{1,2}|^2\int_0^\infty x\biggr[J_{\pm(l+1/2)}(\omega x)\biggl]^2\,dx 
\end{eqnarray}
where $x = 1/\rho$. It is easy to verify that the norm of the scalar modes for the
extreme case near the horizon is divergent. In principle, this problem could be circumvented.
Plane wave solutions in the Minkowski space-time are also divergent when integrated in all space. However, this
difficulty is solved by defining the plane wave modes in a finite volume; at
the end, when the physical quantities are evaluated, the limit of an infinite volume
may be applied. Here, such "normalization" procedure can not be implemented due to
one particularity of the space-time already pointed out: the spatial distance of any
point to the event horizon is infinite; hence, any volume around the horizon
is infinite. It is not possible to normalize the modes found above.
\par
The conclusion is that it is not possible to assign a
complete basis of orthonormal modes near the horizon for the extreme RN black hole.
In this sense, we can understand the negative result of the preceding section:
the global propagation of quantum fields from past infinity to future infinity is
ill definite due to the anomalous behaviour of these quantum fields near the horizon.
The modes cross the shell just before the horizon formation will propagate in
the geometry described before. They are, in this situation, anymore normal modes.
Hence, the basis at the future infinity is not anymore a complete basis
of normal modes.
\par
It is instructive to compare this situation with a similar one that occurs with
AdS black holes \cite{witten,horowitz1}, which plays a crucial r\^ole in
the AdS/CFT correspondence. In this case, the metric takes the form (\ref{metric4}) at
the spatial infinity, corresponding to $\rho \rightarrow \infty$, and not
at $\rho \rightarrow 0$ as in the present case. Again, the solution of the Klein-Gordon
equation takes the form (\ref{kg1}). Now, one of the modes is divergent in the sense of the Klein-Gordon inner product,
while the other one is finite, since the solutions are valid in
the limit $x \rightarrow 0$ in expression (\ref{inner}). The choice of appropriate boundary conditions
allows now to select just the normalizable modes. These features have already been
discussed in references \cite{witten,douglas}. Notice that in this AdS black hole case,
the limit $\rho \rightarrow \infty$ corresponds to the timelike boundary
(where, following \cite{isham}, the choice of convenient boundary conditions
allows to give sense to the Hilbert space in an anti-deSitter space-time), in contrast what
happens with the near horizon limit of the extreme black hole. Again in contrast with
the extreme black hole case, in the AdS black hole the spatial distance of any point
to the horizon is finite.
Finally, it is important to remark that in the extreme black hole geometry, the other
boundary corresponds to a Minkowski space-time, where we find plane wave modes; hence,
there is apparently no way to select appropriate boundary conditions in order to avoid divergent
quantum modes.

\section{Conclusions}

The surface gravity of the extreme Reissner-Nordstr\"om black hole
is zero. For this reason, the temperature of the extreme black hole is
believed to be zero. Moreover, if the temperature for the general
Reissner-Nordstr\"om is evaluated and the extreme condition $M = Q$ is imposed in
the final expression,
it results $T = 0$. But, the notion of temperature must be
extracted, for example, from the direct computation of the
relation between the {\it in} and {\it out} through the
Bogolubov's coefficients. In this paper, this evaluation of the
Bogolubov's coefficients was performed by imposing from the
beginning the extreme condition. We have found that the
Bogolubov's coefficients do not satisfy the normalization
condition (\ref{c3}) due, mainly, to the presence of non
convergent integrals in the computation of the modulus of the
coefficient $\alpha_{\omega\omega'}$. Hence, this semi-classical
analysis seems to be ill definite for the extreme black hole.
\par
This is a quite curious result. In general it is believed that the
normalization condition must be satisfied by construction. If the
normalization condition is not satisfied, it means that the
construction itself is ill definite. What is the reason for the
failure of this construction for the extreme black hole? A very important
point to be noticed concerns the fact that, in spite of what we could think,
the extreme RN black hole is not the limit of the non-extreme ones.
This has already been stressed in reference \cite{hawking1}.
This is due to the behaviour of the geometry near horizon. In the extreme case,
in contrast of what happens in the non-extreme situation, the near horizon geometry
is described by a portion of the anti-deSitter space-time.
\par
The formulation of quantum field theory in an anti-deSitter space-time
is not
a trivial problem. In reference \cite{isham} it has been shown that
a Hilbert space can be implemented in an anti-deSitter space-time
if specific boundary conditions are fixed and if the universal covering
of the anti-deSitter space-time is used. This is due to fact that the
anti-deSitter space-time is not globally hyperbolic and its infinity is
timelike. However, the near horizon geometry for extreme
black hole is just a portion of
the anti-deSitter space-time. Computing the normal modes in this
case leads to non-normalized states. In principle, it is not possible in
this case to recover the notion of Hilbert space, due to the absence of
a universal covering of anti-deSitter space-time. 
\par
Following the computations for the extreme and non-extreme RN black holes
we can easily see that the results obtained in the
general case leads, after imposing the limit case $\kappa = 0$, to the expressions that are
consistent with a zero temperature black hole: the $\beta_{\omega\omega'}$ coefficient goes
to zero, but the $\alpha_{\omega\omega'}$ coefficient remains non null. On the other hand,
the limit
case $\kappa = 0$ when applied to the results obtained for the general case does not lead
to the corresponding results obtained by imposing the extreme condition from the begining. Tracing back
where the discrepancy begins to occur, we may verify that both cases loose contact at very
begining, in the relations between $u$ and $v$ modes (\ref{rela}) and (\ref{relb}), for the
general and extreme case, respectively. Notice that the tortoise coordinate
(\ref{torto2}) may be obtained from (\ref{torto1}) in the extreme limit case.
Again, the specificity of the relation between the {\it in} and {\it out} modes in
the extreme case
is due to matching condition imposed between them at the moment of the horizon formation.
This stress again the crucial role played by the geometry near the event horizon.
\par
In general lines, the results reported in this paper confirm, as far
as the standard semi-classical analysis is concerned, those
of references \cite{liberati1,liberati2} in stating that the extreme black
hole does not behave as thermal object. But, it indicates also that, in principle, no semi-classical
analysis at all can be consistently performed for the extreme black hole. 
This is also in agreement with the results of reference \cite{hawking1},
stating the specificity of the extreme black hole and its failure to obey
the general black holes's thermodynamics laws. Remark that the extreme black holes would be
an example of violation of the third law of thermodynamics.

{\bf Acknowledgement}: We thank G.E. Matsas, G. Cl\'ement and Ph. Spindel for many helpful discussions
on this problem. The remarks made by the anonymous referee were
important in order to improve the final version of this paper.
This work has been partially supported by CNPq (Brazil) and CAPES (Brazil).

\end{document}